\documentclass[aps,prb,epsf,twocolumn,showpacs]{revtex4-2}
\usepackage{CJK,amsmath,amssymb,graphics,epsfig,epstopdf,color,verbatim,ulem,braket,tabularx,float}
\usepackage[colorlinks,linkcolor=blue,citecolor=blue,urlcolor=blue]{hyperref}

\begin{document}

\title{Exact diagonalization study of triangular Heisenberg model with four-spin ring-exchange interaction}

\author{Yuchao Zheng}
\affiliation{Guangdong Fundamental Research Center for Magnetoelectric Physics, \\Guangdong Provincial Key Laboratory of Magnetoelectric Physics and Devices,\\Center for Neutron Science and Technology, School of Physics, Sun Yat-sen University, Guangzhou 510275, China}

\author{Muwei Wu}
\affiliation{Guangdong Fundamental Research Center for Magnetoelectric Physics, \\Guangdong Provincial Key Laboratory of Magnetoelectric Physics and Devices,\\Center for Neutron Science and Technology, School of Physics, Sun Yat-sen University, Guangzhou 510275, China}

\author{Dao-Xin Yao}
\email{yaodaox@mail.sysu.edu.cn}
\affiliation{Guangdong Fundamental Research Center for Magnetoelectric Physics, \\Guangdong Provincial Key Laboratory of Magnetoelectric Physics and Devices,\\Center for Neutron Science and Technology, School of Physics, Sun Yat-sen University, Guangzhou 510275, China}

\author{Han-Qing Wu}
\email{wuhanq3@mail.sysu.edu.cn}
\affiliation{Guangdong Fundamental Research Center for Magnetoelectric Physics, \\Guangdong Provincial Key Laboratory of Magnetoelectric Physics and Devices,\\Center for Neutron Science and Technology, School of Physics, Sun Yat-sen University, Guangzhou 510275, China}

\begin{abstract}

Using Lanczos exact diagonalization (ED), we study the spin-1/2 $J_1$-$J_2$ Heisenberg model with the four-spin ring-exchange interaction $J_r$ on triangular lattice. We mainly use the level spectroscopic technique of two 36-site tori to investigate the ground-state phase diagram, and further characterize phases by spin, dimer and chiral correlation functions. The ground state has rich phases including several magnetic ordered phases like zigzag phase and tetrahedral phase, as well as several novel nonmagnetic phases, some of which exhibit valence bond solid behavior in their dimer correlation functions. However, we do not find direct evidence of a quantum spin liquid phase with spinon Fermi surface in this model. Our results can give a better understanding of the ground-state properties of the triangular Heisenberg model with ring-exchange interaction, and help to understand the relevant triangular materials.

\end{abstract}

\date{\today}
\maketitle

\section{Introduction}

Frustrated system can present novel states and phenomena such as complex magnetism and quantum spin liquid (QSL)~\cite{Anderson1973SL,Balents2010SL}. A correct understanding of these new phenomena is of great significance to the study of strongly correlated electron systems. In triangular Heisenberg model, the presence of frustration, low spin, next-nearest neighbors, or even more complex interactions may lead to exotic quantum phases and unconventional phase transitions. In recent years, the possible QSLs and novel excitations in triangular lattice have attracted considerable attention. In experiments, spin-liquid-like behaviors have been observed in many spin-1/2 antiferromagnetic materials with triangular lattice, including $\kappa$-(ET)$_2$Cu$_2$(CN)$_3$~\cite{Shimizu2003ET2Cu2CN3,Kurosaki2005ET2Cu2CN3,Yamashita2008ET2Cu2CN3}, EtMe$_3$Sb[Pd(dmit)$_2$]$_2$~\cite{Itou2007EtMe3SbPddmit22JOP,Itou2007EtMe3SbPddmit22PRB,Yamashita2010EtMe3SbPddmit22,Yamashita2011EtMe3SbPddmit22,Bourgeois-Hope2019EtMe3SbPddmit22}, YbMgGaO$_4$~\cite{Li2015YbMgGaO4PRL,Li2015YbMgGaO4SR,Li2016YbMgGaO4,Shen2016YbMgGaO4,Xu2016YbMgGaO4,Paddison2017YbMgGaO4,Wu2021YbMgGaO4}, 1$T$-TaS$_2$~\cite{Klanjvsek20171T-TaS2,Law20171T-TaS2,Ribak20171T-TaS2,Murayama20201T-TaS2}, NaYbO$_2$~\cite{Liu2018NaYbO2,Bordelon2019NaYbO2,Lei2019NaYbO2} and Na$_2$BaCo(PO$_4$)$_2$~\cite{Zhong2019Na2BaCoPO42,xiang2024Na2BaCoPO42}. However, in most of the triangular materials mentioned above, there are still some controversies about the type of QSL and the existence of a QSL phase with a spinon Fermi surface (SFS).

Theorists have made significant efforts to study different types of triangular Heisenberg models and have tried to predict exotic phases and resolve experimental controversies. Up to now, we have known that the ground state of the pure isotropic Heisenberg model on a triangular lattice is the 120$^{\circ}$ antiferromagnetic (AFM) state~\cite{Bernu1994Tri120AFM,Capriotti1999Tri120AFM,White2007Tri120AFM}. Adding the next-nearest-neighbor interaction $J_2$ or spin-orbital coupling terms allow the system to enter a small regional QSL phase~\cite{Kaneko2014J1J2Tir,Hu2015J1J2Tri,Li2015J1J2Tri,Zhu2015J1J2Tri,Iqbal2016J1J2TriDSL,Saadatmand2016J1J2Tri,Ferrari2019J1J2Tri,Hu2019J1J2TriDSL,Tang2022J1J2TriExitedStateCross,Li2015YbMgGaO4PRL,Li2016J1J2Tri,Liu2016J1J2Tri,Luo2017J1J2Tri,Zhu2017J1J2Tri,Maksimov2019J1J2Tri,Wu2019J1J2Tri,Wu2021YbMgGaO4,Lauchili2024J1J2Tri}, which is more likely to be considered as a Dirac spin liquid~\cite{Iqbal2016J1J2TriDSL,Hu2019J1J2TriDSL}, but this is not consistent with some experimental measurements on SFS-QSL candidate materials, like $\kappa$-(ET)$_2$Cu$_2$(CN)$_3$, EtMe$_3$Sb[Pd(dmit)$_2$]$_2$, 1$T$-TaS$_2$. In their heat capacity measurements, a linear T term was observed, suggesting the existence of SFS~\cite{Yamashita2008ET2Cu2CN3,Yamashita2010EtMe3SbPddmit22,Ribak20171T-TaS2,Law20171T-TaS2,Murayama20201T-TaS2}. A nonzero value of the thermal conductivity in the zero-temperature limit was observed in EtMe$_3$Sb[Pd(dmit)$_2$]$_2$, 1$T$-TaS$_2$~\cite{Yamashita2010EtMe3SbPddmit22,Murayama20201T-TaS2}, which also signifies the existence of SFS. To better describe these triangular materials that are in the close vicinity of the Mott transition and to realize the SFS-QSL phase in a theoretical model, a four-spin ring exchange interaction $J_r$ induced by charge fluctuations has been proposed~\cite{Motrunich2005RingVS}. Incorporating $J_r$ can lead to a rich variety of phenomena~\cite{Thouless1965He3,Li2023ringTripleQ,Liu2023ringTripleQ,Ma2024TripleQ,Dai2017ColdAtom}. Different numerical methods have been employed for calculations, but the results are still a subject of debate~\cite{Motrunich2005RingVS,Mishmash2013RingVMC,Xu2018RingDMRG,Seki2020ringED,Liu2021RingVMC,Cookmeyer2021RingEDDMRG,LiJX2023RingVMC,LiT2023RingVSED,Schultz2024RingDMRG,Mitra2024RingDMRG}. In earlier studies, variational study~\cite{Motrunich2005RingVS} found that upon including the ring exchanges, the 120° AFM state gives way to the SFS state. A few years later, several variational Monte Carlo (VMC) studies on the $J_1$-$J_2$-$J_r$ model~\cite{Mishmash2013RingVMC,Liu2021RingVMC,LiJX2023RingVMC} consistently found that the SFS state becomes the best variational ground state when $J_r$ is relatively large. However, a recent VMC study~\cite{LiT2023RingVSED} suggested that such a state is never favored in this model on the triangular lattice. Density matrix renormalization group (DMRG) simulations have also led to different conclusions: for $J_r \geq 0.4$, an earlier DMRG study~\cite{Xu2018RingDMRG} considered the SFS state to be the ground state, while more recent DMRG studies~\cite{Cookmeyer2021RingEDDMRG,Schultz2024RingDMRG,Mitra2024RingDMRG} found a zigzag order state in the same region. Therefore, there are still controversies regarding whether the SFS state exists on this kind of model\cite{Motrunich2005RingVS,Mishmash2013RingVMC,Xu2018RingDMRG,Liu2021RingVMC,Cookmeyer2021RingEDDMRG,LiJX2023RingVMC,LiT2023RingVSED,Schultz2024RingDMRG,Mitra2024RingDMRG}. 

In this work, we employ exact diagonalization to reinvestigate the spin-1/2 $J_1$-$J_2$-$J_r$ Heisenberg model on triangular lattice and detect the possible exotic phases and phase transitions. In the ED calculations, we mainly use energy level crossings of two 36-site tori to obtain the approximate phase boundaries. And then use correlation functions to further characterize these phases. Even though 36-site tori have some finite-size effects, it is crucial for us to understand the ground-state phases. Our findings reveal a rich variety of phases within the model, with several nonmagnetic phases exhibiting valence bond solid (VBS)-like characteristics. Most significantly, our results do not corroborate the presence of the SFS-QSL phase in this model.  Our results are very helpful to clarify the effects of four-spin ring exchange in the triangular model. 

This paper is organized as follows. In Sec.~\ref{sec:Model and method}, we introduce the $J_1$-$J_2$-$J_r$ model on triangular lattice and the exact diagonalization numerical methods. In Sec.~\ref{sec:J1Jr phase diagram} and Sec.~\ref{sec:J1J2Jr phase diagram}, we present detailed descriptions of the ground-state phase diagram and investigate properties of each phases. We also compare our results with previous studies. Section~\ref{sec:Conclusion} will give a brief summary.

\section{Model and method}\label{sec:Model and method}

\begin{figure}[t]
	\centering
	\includegraphics[width=\linewidth]{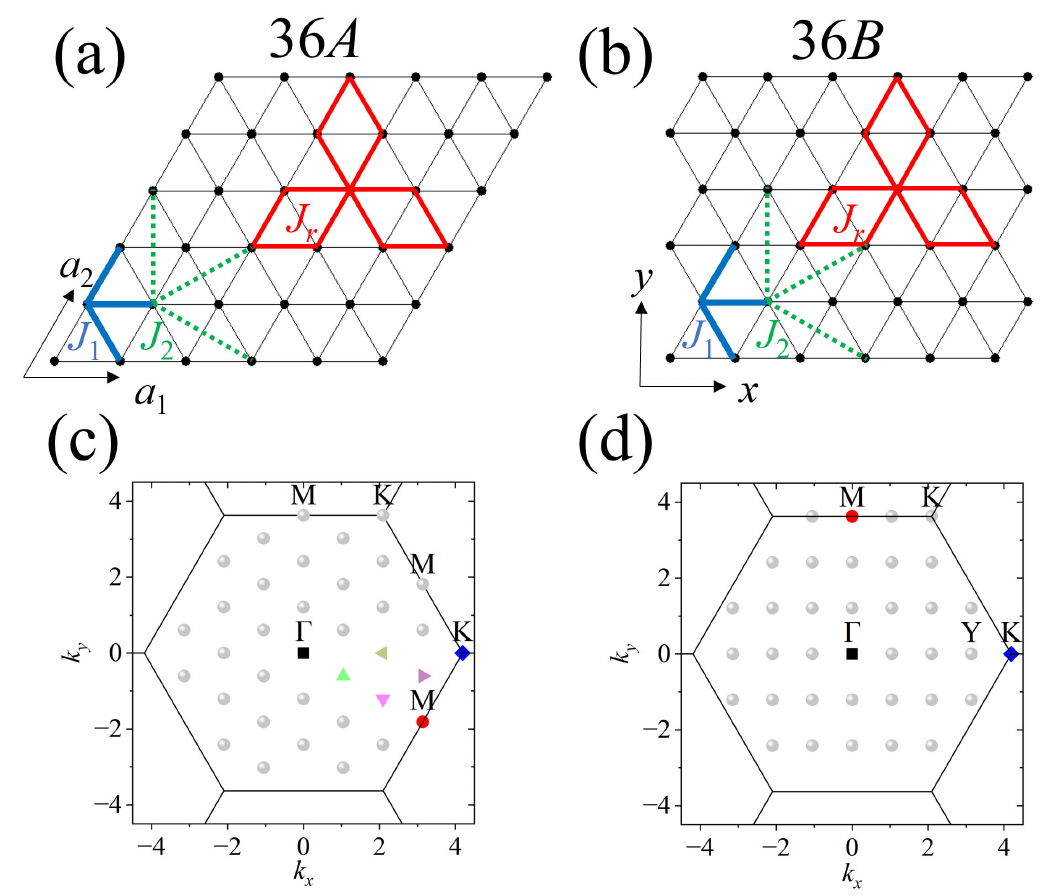}
        \caption{%
		(a) and (b) are two 36-site clusters (or tori) with two different periodic boundary conditions (PBC), which are called 36$A$ and 36$B$ respectively. The black arrows at the bottom left represent the PBC directions. Different colored lines represent the corresponding interactions. (c) and (d) are the first Brillouin zones of the 36$A$ and 36$B$ clusters respectively. Solid lines denote the Brillouin-zone boundaries. Symbols indicate the 36 momenta inside the first Brillouin zone. Colorful symbols are the momenta used in the ED calculations.
        }
        \label{fig:LattBZ}
\end{figure}

The Hamiltonian we consider here read as following:
\begin{align}
H=&J_1\sum\limits_{\left\langle {i,j} \right\rangle }\mathbf{S}_i\cdot\mathbf{S}_j+J_2\sum\limits_{\left\langle {\left\langle {i,j} \right\rangle } \right\rangle }\mathbf{S}_i\cdot\mathbf{S}_j\notag\\
&+J_r\sum\limits_{\left\langle {i, j, k, l} \right\rangle }[(\mathbf{S}_i\cdot\mathbf{S}_j)(\mathbf{S}_k\cdot\mathbf{S}_l)
+(\mathbf{S}_i\cdot\mathbf{S}_l)(\mathbf{S}_j\cdot\mathbf{S}_k)\notag\\
&-(\mathbf{S}_i\cdot\mathbf{S}_k)(\mathbf{S}_j\cdot\mathbf{S}_l)],
\label{eq:Hmlt}
\end{align}
in which we consider the nearest-neighbor interaction $J_1$, the next-nearest-neighbor interaction $J_2$ and the four-spin ring-exchange interaction $J_r$. ${\left\langle {i,j} \right\rangle }$ and ${\left\langle\left\langle {i, j} \right\rangle\right\rangle }$ denotes nearest, next-nearest neighbor pairs, respectively. ${\left\langle {i, j, k, l} \right\rangle }$ denotes a sum over all the minimal rhombuses [see Figs.~\ref{fig:LattBZ}(a) and~\ref{fig:LattBZ}(b)]. The $J_r$ term can be obtained from the perturbation expansion of the Hubbard model~\cite{Takahashi1977Jr,MacDonald1988JrModel,Delannoy2005JrModel} and can be enhanced when the system gets closer to the Mott transition. In the following, we will set $J_1\equiv 1$ as the energy unit and lattice constant $a\equiv 1$ as the length unit.

In this work, we mainly perform ED calculation on two kinds of 36-site tori including 36$A$ and 36$B$ [see Figs.~\ref{fig:LattBZ}(a) and~\ref{fig:LattBZ}(b)]. 36$A$ torus shares the same $C_{6v}$ point-group symmetry as the infinite lattice. Therefore, this torus geometry is important for determining the phase boundaries using ground state and excited state level crossings~\cite{Wietek2017TOS}. We also compute the correlation functions, including spin correlation, dimer correlation and scalar chiral correlation, to characterize different phases. To reduce the computational cost, we use several symmetries, including $U(1)$ symmetry, translational symmetry, and spin inversion symmetry, to do the Lanczos diagonalization and obtain the low-lying eigenvalues. We restrict our calculations in the $\sum_{i}\langle\mathbf{S}_i^z\rangle=0$ block with the same number of $\uparrow$ and $\downarrow$ spins. The maximum block dimension is about 126 million. More specifically, we only diagonalize the Hamiltonian block with specific momenta shown with different symbols and colors in Figs.~\ref{fig:LattBZ}(c) and~\ref{fig:LattBZ}(d) due to the point-group symmetries. When we presenting the low-lying energy spectrum, we also use solid and hollow symbols to distinguish eigenvalues in different spin-inversion symmetry sectors. To use the translational and spin-inversion symmetry in ED, we construct a set of states,
\begin{equation}
    |a(k, z) \rangle = \frac{1}{\sqrt{N_a}}\sum\limits_{\mathbf{r}}e^{-i\mathbf{k} \cdot \mathbf{r}}T^{\mathbf{r}}(1+zZ) |a \rangle,
\end{equation}
in which $T$ and $Z$ are respectively translation and spin inversion operators, $z$ is the eigenvalue of $Z$, $|a \rangle$ is a reference state with $m_z$ = 0. By taking these states as basis, the Hamiltonian matrix breaks up into some smaller blocks with different momentum $\mathbf{k}$ and $z$. To get the excited states, after convergence of the ground state, we start the Lanczos process to construct a new initial Krylov vector which is orthogonal to all previous low energy states and then target excited states one by one~\cite{Sandvik2010ED,Lauchli2011ED}.

\begin{figure*}[!t]
	\centering
	\includegraphics[width=\linewidth]{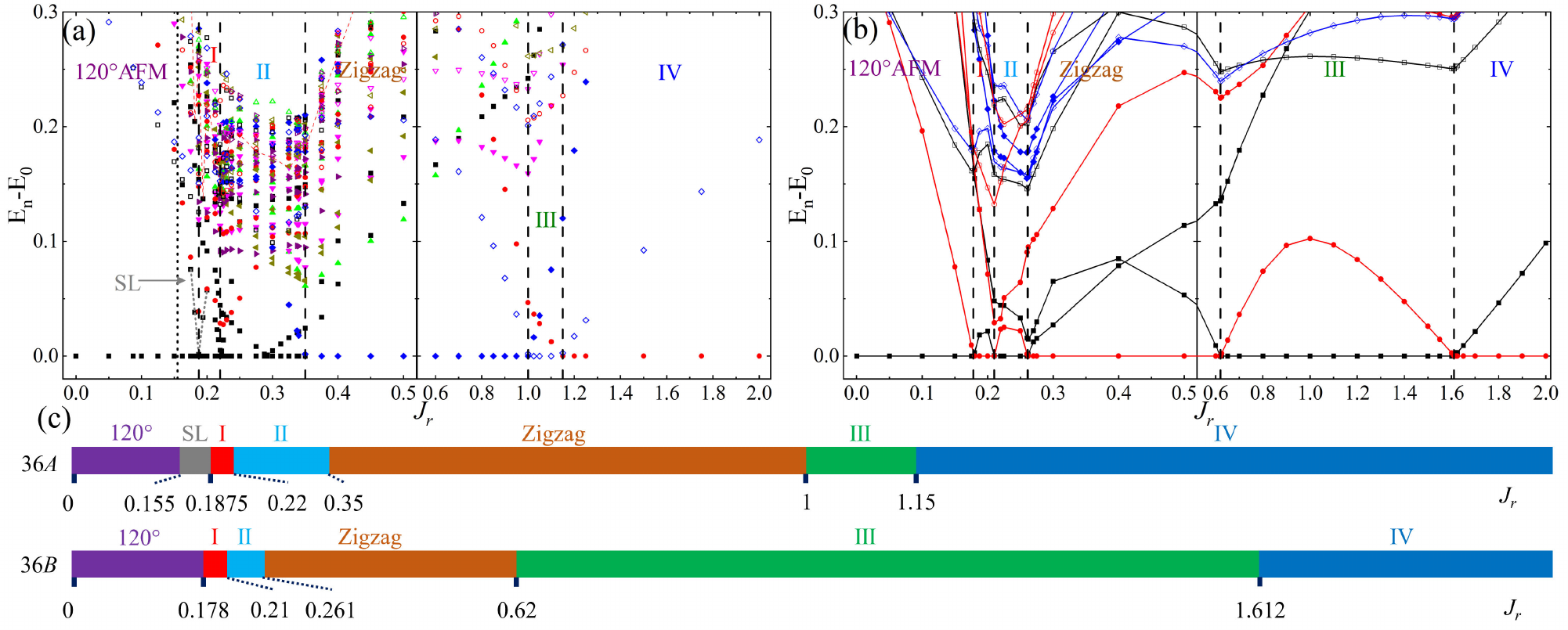}
        \caption{%
		(a) is the energy spectra of 36$A$ as functions of $J_r$ when $J_2=0$. The vertical black dashed lines indicate the phase boundaries. Symbols with different shapes and colors indicate different momenta [refer to Figs.~\ref{fig:LattBZ}(c) and~\ref{fig:LattBZ}(d), the rest momenta with gray color are equivalent to momenta with other colors under point-group symmetries for 36$A$.], the solid and hollow symbols indicate different spin-inversion sectors. The energies below the red dashed line form the full lowest spectra without missing any energy levels. (b) is the energy spectra of 36$B$ as functions of $J_r$ when $J_2=0$. Compared to 36$A$, for 36$B$, we only present two energy levels at each high-symmetry momentum sector ($\Gamma,K,M$). We have also calculated energy levels at other non-equivalent momenta (not shown in the figure) to ensure that the ground state energy is accurate and that there are no other level crossings between the ground state and the excited states. (c) $J_1-J_r$ phase diagram obtained on 36$A$ and 36$B$. We can identify similar phases using these two tori, but the phase boundaries are quite different for some of these phases.}
        \label{fig:EngySpectrumJP0p0}
\end{figure*}

To detect the possible conventional ordering like magnetic order, dimer order, and chiral order, we compute the corresponding correlation functions. The structure factor of spin correlation $S(\mathbf{q})$ is defined as:
\begin{equation}
S(\mathbf{q})=\frac{1}{N}\sum_{i,j}e^{-i\mathbf{q}\mathbf{r}_{ij}}\left\langle {\mathbf{S}_i\cdot\mathbf{S}_j} \right\rangle,
\end{equation}
the structure factor of dimer correlation $D(\mathbf{q})$ is defined as:
\begin{eqnarray}
D(\mathbf{q})&=&\frac{1}{3N}\sum_{\alpha}\sum_{i,j}e^{-i\mathbf{q}(\mathbf{r}_i-\mathbf{r}_j)}\left\langle {\hat{\mathbf{B}}_{i\alpha}\hat{\mathbf{B}}_{j\alpha}} \right\rangle,\nonumber\\
\hat{\mathbf{B}}_{i_\alpha}&=&\mathbf{S}_i\cdot\mathbf{S}_{i+\alpha}-\left\langle {\mathbf{S}_i\cdot\mathbf{S}_{i+\alpha}} \right\rangle, \label{eq:Dq-alpha}
\end{eqnarray}
where $i+\alpha$ represents one of the three nearest neighbor positions of $i$, $\mathbf{r}_i$ represents the vector of position $i$, and the structure factor of chiral correlation $\chi(\mathbf{q})$ is defined as:
\begin{eqnarray}
\chi(\mathbf{q})&=&\frac{1}{N_{\triangle}}\sum_{i,j\in \triangle}e^{-i\mathbf{q}(\mathbf{r}_i-\mathbf{r}_j)}\left\langle {\hat{\mathbf{\chi}}_i\hat{\mathbf{\chi}}_j} \right\rangle,\nonumber\\
\hat{\mathbf{\chi}}_i&=&\mathbf{S}_a\cdot(\mathbf{S}_{b}\times\mathbf{S}_{c}),
\end{eqnarray}
where $\mathbf{r}_i$ denotes the position vector to the center of the upward-pointing triangle $i$, and $a,b$ and $c$ are the sites (vertices) of the triangle $i$.

\section{\texorpdfstring{$J_2 = 0$, $J_1$-$J_r$ phase diagram}{}}\label{sec:J1Jr phase diagram}

\subsection{\texorpdfstring{$0 < J_r < 1$}{}}

\begin{figure*}[t!]
	\centering
	\includegraphics[width=\textwidth]{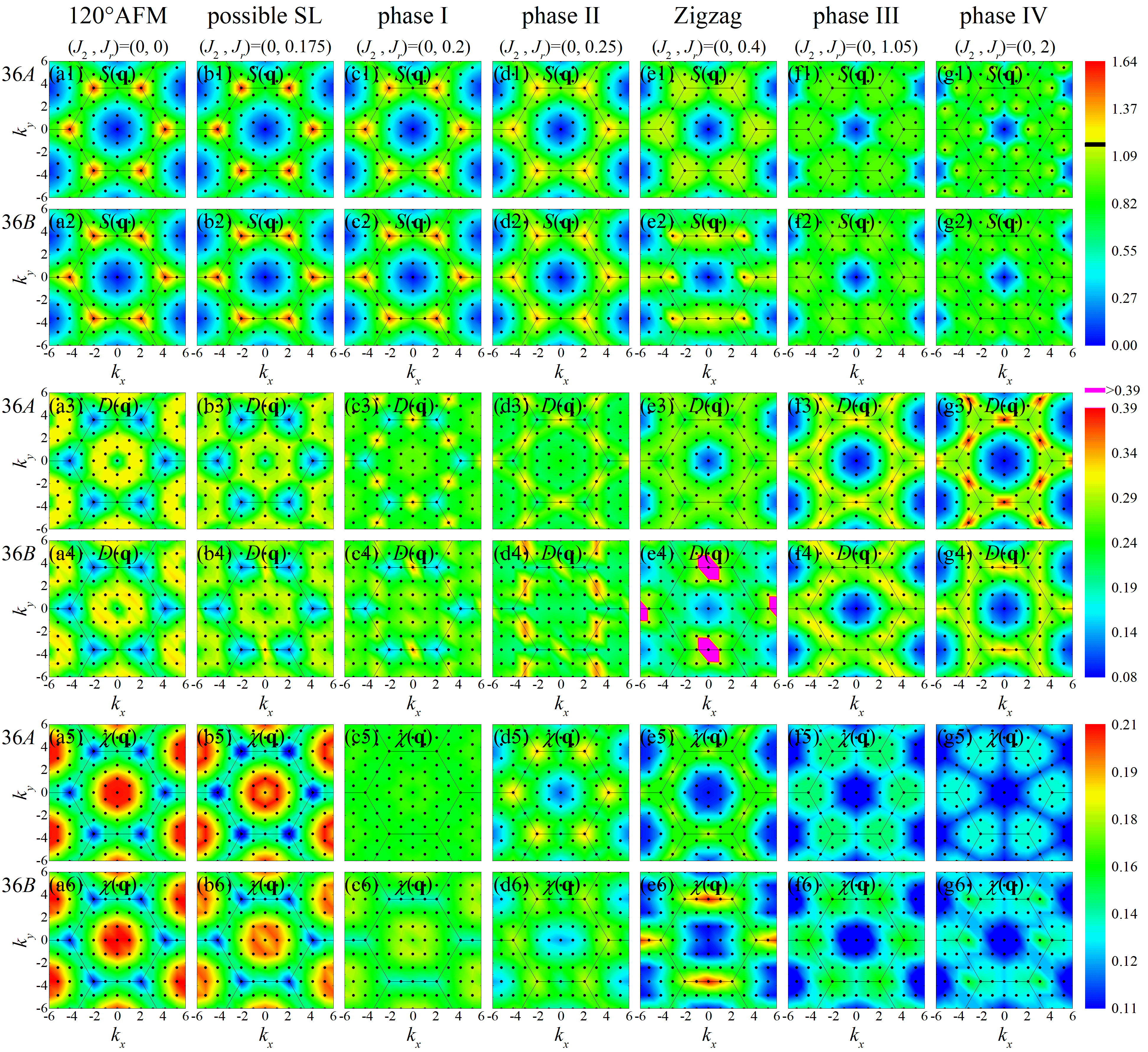}
        \caption{%
		Structure factors for $120^{\circ}$ AFM phase, possible SL phase, phase I, phase II, zigzag phase, phase III and phase IV, respectively. Each column of figures corresponds to the different parameters (phases). The top two rows represent spin structure factors. The first row corresponds to the results obtained on 36$A$, and the second row corresponds to the results obtained on 36$B$. Similarly, the middle two rows show the dimer structure factors, and the last two rows show the chiral structure factors, with the first row in each pair corresponding to 36$A$ and the second row to 36$B$.
        For spin structure factors, above the boundary value $U_0 = 1.13$ labeled by a black line on the color bar, a logarithmic mapping is used, $U = U_0 + \log_{10} [S(\mathbf{q})] -\log_{10} [U_0]$, and $U = S(\mathbf{q})$ while below the boundary value.
        }
        \label{fig:Corr-K}
\end{figure*}

\begin{figure*}[t!]
	\centering
	\includegraphics[width=\textwidth]{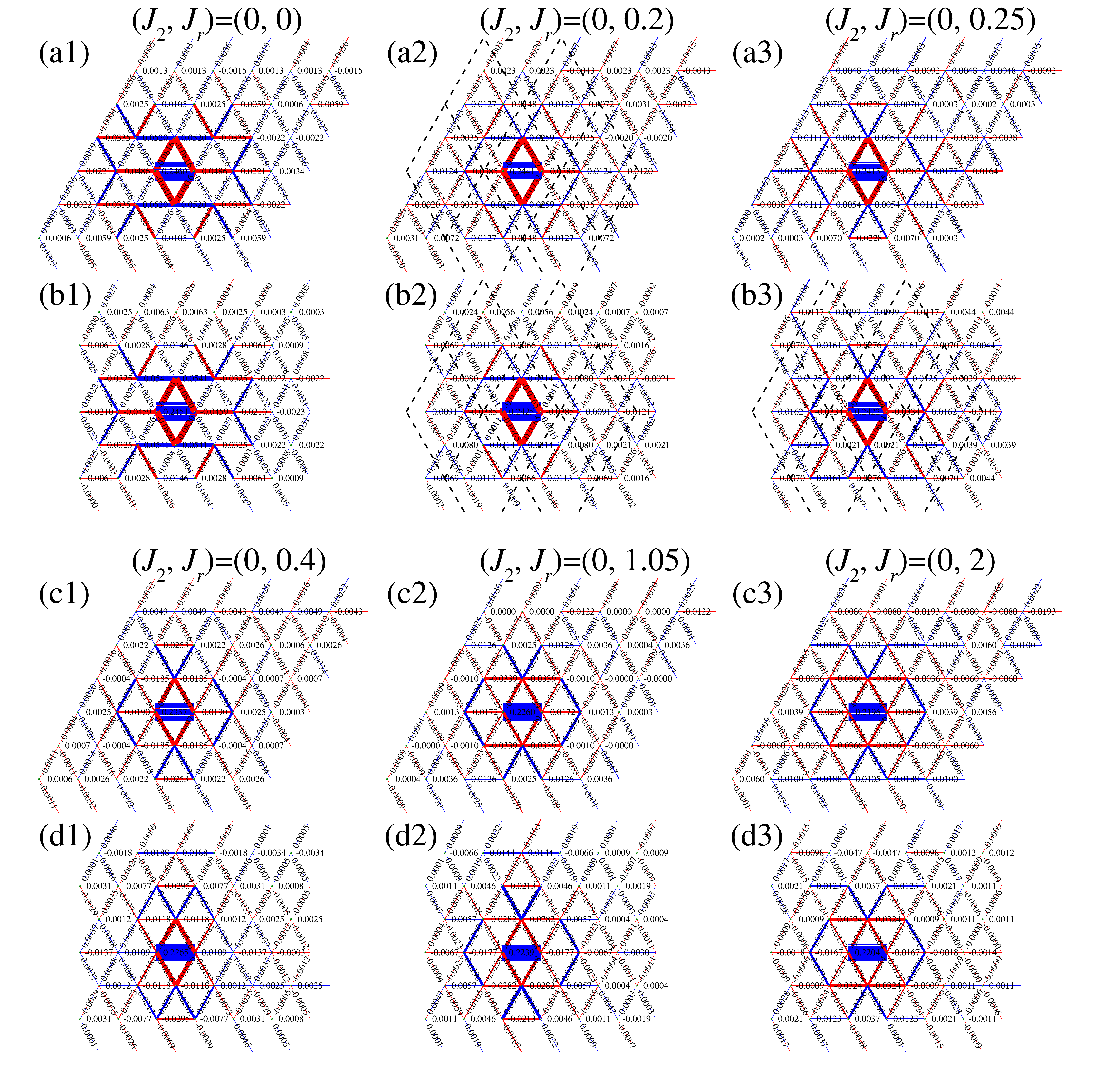}
        \caption{%
		(a1)-(a3) and (b1)-(b3) show the real-space dimer correlations for $120^{\circ}$ AFM phase, phase I and phase II in 36$A$ and 36$B$, respectively. (c1)-(c3) and (d1)-(d4) show the real-space dimer correlations for zigzag phase, phase III and IV in 36$A$ and 36$B$, respectively. The corresponding $J_2$ and $J_r$ are marked at the top of corresponding figures. In each figure, the thickest horizontal blue bond in the middle represents the reference bond. Bond thickness corresponds to the magnitude of dimer correlation, while color indicates the sign: blue signifies positive, and red signifies negative. The black dash rhombus frames indicate the unit cell of possible VBS pattern.
 }
        \label{fig:Corr-R}
\end{figure*}

We first consider the $J_1$-$J_r$ model with $J_2=0$. Figure~\ref{fig:EngySpectrumJP0p0} shows the energy spectra and phase diagrams of the $J_1$-$J_r$ model as functions of $J_r$, obtained from two 36-site tori. At $J_r=0$, the system is in the 120$^{\circ}$ AFM phase. This magnetic ordering breaks the $SU(2)$ symmetry in the thermodynamic limit, leading to the famous Anderson tower of states in finite-size systems~\cite{Anderson1952TOS,Claire2005TOS,Wietek2017TOS}. 
In the energy spectra of the 36$A$ torus, the ground state of the 120$^{\circ}$ AFM phase is a singlet with momentum $\Gamma$ and even spin-inversion parity with spin-inversion eigenvalue $+1$ [labeled as ($\Gamma, 1$)], and the lowest three excited states are triplets with odd spin-inversion parity [one ($\Gamma, -1$) and two ($K, -1$)]. These quantum numbers correspond to a set of states that exhibit $SU(2)$ symmetry, as analyzed in reference~\cite{Wietek2017TOS}. The structure factor of spin correlation $S(\mathbf{q})$ has strong peaks at $K$ points in this phase [see Fig.~\ref{fig:Corr-K}(a1)], while the dimer and chiral correlation functions show no obvious long-range order, indicating the 120$^{\circ}$ AFM ordering. At $J_r\approx 0.155$, the two-fold degenerate lowest excited triplet ($K, -1$) intersects with the two-fold degenerate lowest excited singlet ($\Gamma, 1$), highlighted with a vertical black dotted line at $J_r\approx 0.155$ in Fig.~\ref{fig:EngySpectrumJP0p0}(a). We consider the intersection of these first excited states from different symmetry sectors as indication of a phase transition to a possible spin liquid (SL) phase~\cite{Tang2022J1J2TriExitedStateCross,Sandvik2016ExitedStateCross,Sandvik2018ExitedStateCross,Nomura2021ExitedStateCross,Sandvik2022ExitedStateCross,Lauchili2024J1J2Tri} which adiabatically connects to the Dirac spin liquid phase at the $J_1$-$J_2$ model. In Appendix~\ref{App:J1J2}, we also show the similar crossing between excited states in the $J_1$-$J_2$ model, which can be used to determine the quantum critical point between the 120$^{\circ}$ AFM phase and the SL phase, and which is exactly the same as the energy spectra shown in Ref.~\onlinecite{Lauchili2024J1J2Tri}.
The adiabatic connection of SL phase in two regions becomes more obvious when considering the next-nearest-neighbor $J_2$ interaction. We will provide more discussions in the following section.

As $J_r$ increases, the ground states and the first-excited states cross at $J_r\approx0.1875, 0.22, 0.29, 0.35$ and $1.0$ in the 36$A$ torus, highlighted with vertical black dashed lines in Fig.~\ref{fig:EngySpectrumJP0p0}(a). These imply the possible first-order phase transitions in the thermodynamic limit. It should be noted that $J_r\approx0.29$ is not considered as a phase transition point, because their correlation functions are similar without abrupt change. For the 36$B$ torus, a similar phase diagram can be obtained by examining the level crossings between the ground states and the excited states [we only present two energy levels at each high-symmetry momentum sectors as shown in Fig.~\ref{fig:EngySpectrumJP0p0}(b)]. However, the phase boundaries differ due to finite-size effects. We will focus more on the 36$A$ torus because it respects the full symmetry of the entire lattice. Furthermore, except for the zigzag phase, it seems that 36$A$ is better at finding characteristic peaks in the momentum space correlation functions (see Fig.~\ref{fig:Corr-K}).

During the transition from the SL phase to phase I, indicated by the vertical dashed line at $J_r\approx 0.1875$ in Fig.~\ref{fig:EngySpectrumJP0p0}(a), a two-fold degenerate singlet state ($\Gamma, 1$) intersects with the previous one-fold ground-state singlet ($\Gamma, 1$) [highlighted by a "V"-shape gray dotted line around $J_r\approx0.1875$ in Fig.~\ref{fig:EngySpectrumJP0p0}(a)], and becomes the ground state within phase I. The structure factors of phase I are presented in Figs.~\ref{fig:Corr-K}(c1-c6). Within this phase, the structure factor $S(\mathbf{q})$ exhibits weaker peaks at the $K$ points compared to those in the 120$^{\circ}$ AFM phase, indicating the possible absence of magnetic ordering in the thermodynamic limit. Unlike the 120$^{\circ}$ AFM phase, the dimer structure factor $D(\mathbf{q})$ exhibits peaks at the $M$ points, suggesting the existence of a possible valence bond solid (VBS)-like phase. The real-space dimer correlations depicted in Figs.~\ref{fig:Corr-R} (a2) and~\ref{fig:Corr-R}(b2) reveal a rhombic dimer pattern with a $2\times2$ periodicity. However, this pattern could also represent the superposition of collinear VBS, where singlets form along specific collinear pattern. Nevertheless, a conclusive confirmation of the existence of this long-range VBS-like phase is beyond 36-site tori.

At $J_r\approx0.22$, the ground state turns from phase I to phase II. Within this phase, another singlet state ($\Gamma, 1$) from high energy transitioning to become the new ground state. 
The structure factor $S(\mathbf{q})$ in this phase exhibits diminished peaks at $K$ points compared to phase I. The peaks of dimer structure factor $D(\mathbf{q})$ at the $M$ points suggest a potential relation to the VBS-like phase. 
From the real-space dimer correlation in Figs.~\ref{fig:Corr-R}(a3) and~\ref{fig:Corr-R}(b3), we observe that the rhombic dimer patterns vanish in 36$A$ torus, but still remain in 36$B$ torus. Furthermore, the chiral structure factor $\chi(\mathbf{q})$ exhibits broad peaks at the $K$ points, a feature that distinguishes it from phase I where the broad peaks are at a circular ring around $\Gamma$ point. At $J_r\approx0.29$, two singlet states cross with each other. However, within the range $0.22<J_r<0.35$, we observe no abrupt changes in the correlation functions. The persistence of structure factor peaks at the same momenta suggests that this regime is in the same phase (see Appendix.~\ref{App:phase2}).  

In the regime where $0.35<J_r<1$, all structure factors, including spin, dimer and chiral, show no signs of ordering in the 36$A$ torus. In this phase, the ground state exhibits a total four-fold degenerate singlet at two K points with spin-inversion eigenvalue $z$=+1. The structure factor $S(\mathbf{q})$ within this phase exhibits broad peaks in the 36$A$ torus [see Fig.~\ref{fig:Corr-K}(e1)], which is attributed to the absence of certain wave vector points that would show magnetic Bragg peaks. To gain further insight, we computed the 36$B$ torus and observed clear peaks of $S(\mathbf{q})$ at the $Y(\pi, 0)$ points in the Brillouin zone, as seen in Fig.~\ref{fig:Corr-K}(e2). The spin correlation in real space suggests the presence of a zigzag-type magnetic order, where spins align antiferromagnetically along one of the three zigzag directions~\cite{Cookmeyer2021RingEDDMRG}. The 36$A$ torus does not match this kind of magnetic order. And the sharp peak of the dimer structure factor using 36$B$ torus [$D(M)=0.996$ when $J_r=0.4$] shown in Fig.~\ref{fig:Corr-K}(e4) is a manifestation of the magnetic zigzag phase. Similar behavior has also been seen in the collinear AFM phase of $J_1$--$J_2$ Heisenberg model on square lattice shown in Appendix.~\ref{App:SquareJ1J2}.

The $J_1$-$J_r$ phase diagram obtained from the 36$A$ torus is consistent with previous infinite Density Matrix Renormalization Group (iDMRG) results in Ref.~\onlinecite{Cookmeyer2021RingEDDMRG} and Ref.~\onlinecite{Schultz2024RingDMRG}, even though the phase boundaries have some differences and we identify one more phase in the $0 < J_r < 0.4$ region. Between the 120$^\circ$ AFM phase and the zigzag phase, we identify the SL phase, phase I, and phase II, while they identified the CSL and VBS phases. In their papers, the parameter regime similar to our phase I is identified as a chiral spin liquid through the entanglement spectrum which shows the degeneracy pattern expected for the Kalmeyer-Laughlin wavefunction. In our calculations, the SL phase is adiabatically connected to QSL in the $J_1-J_2$ model. And in the phase I, the scalar chiral correlation is weak compared to 120$^{\circ}$ AFM phase. However, the dimer correlation is more clearly to show sharp peaks, and its real-space correlation shows rhombic dimer pattern with $2\times2$ periodicity. It is more like a collinear or plaquette valence bond solid (VBS) phase. For the similar regime of phase II, previous iDMRG studies suggest that this is a collinear VBS phase. Our results also indicate it may be a VBS phase. But we cannot see a clear pattern in real-space dimer correlation using the 36$A$ torus. In 36$B$ torus, we can see the rhombic dimer pattern in both phase I and phase II. This controversy needs further study. Earlier, Ref.~\onlinecite{Xu2018RingDMRG} and Ref.~\onlinecite{Liu2021RingVMC} found a $U(1)$ quantum spin liquid (QSL) with spinon Fermi surface (SFS) when \(J_r > 0.3\). Currently, we find a zigzag phase in this regime and do not find a direct evidence to see the SFS-QSL in our ED results.

\subsection{\texorpdfstring{$J_r > 1$}{}}

We have extended the study to large $J_r$ and find two new phases, namely phase III and phase IV.

In phase III where $1<J_r<1.15$, the ground state exhibits a two-fold degenerate triplet ($K, -1$) on 36$A$ torus. As depicted in Figs.~\ref{fig:Corr-K}(f1-f6), we observe very weak and broad peaks in $S(\mathbf{q})$, $D(\mathbf{q})$ and $\chi(\mathbf{q})$, indicating the absence of conventional order within this phase. This observation suggests the formation of a nonmagnetic phase. What this phase is and whether this phase is a SL are still open questions. 

 When $J_r>1.15$, the system enters phase IV. In this phase, the ground state is a three-fold degenerate ($M, 1$) state. The peaks of $S(\mathbf{q})$ are relatively weak. Additionally, the chiral structure factor $\chi(\mathbf{q})$ does not exhibit significant features. However, $D(\mathbf{q})$ exhibits pronounced peaks at the $M$ point, suggesting a possible VBS phase. The real-space dimer correlation, as depicted in Fig.~\ref{fig:Corr-R}(c3), reveals a clear stripe dimer pattern in 36$A$ which is a strong signature of a VBS phase. Refs.~\onlinecite{Motrunich2005RingVS,Mishmash2013RingVMC,Liu2021RingVMC,LiJX2023RingVMC, Xu2018RingDMRG} have found that the SFS-QSL emerges when \(J_r\) is large, However, we do not find direct evidence of this phase at large $J_r$. Further research is needed to elucidate the precise nature of all the nonmagnetic phases in the $0 < J_r < 2.0$ regime.

\section{\texorpdfstring{$J_1$-$J_2$-$J_r$ phase diagram}{}}\label{sec:J1J2Jr phase diagram}

\begin{figure}[t]
	\centering
	\includegraphics[width=0.9\linewidth]{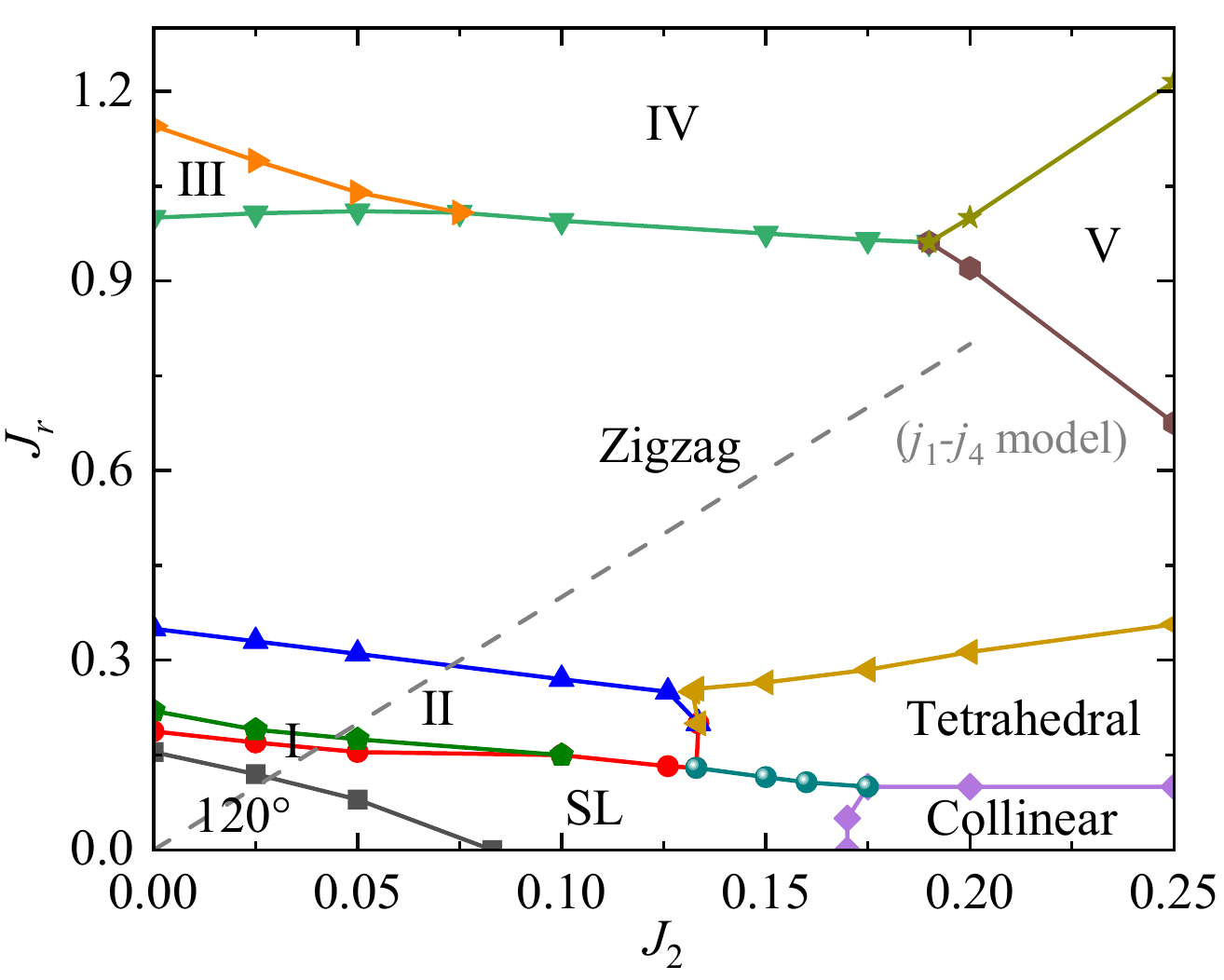}
        \caption{%
        The ground state phase diagram of the $J_1$-$J_2$-$J_r$ model obtained from the 36$A$ torus. The gray dashed line indicates the $j_1-j_4$ model used in some variational Monte Carlo calculations~\cite{Mishmash2013RingVMC,LiJX2023RingVMC,LiT2023RingVSED}.
		}
        \label{fig:PhaseDiagram}
\end{figure}

In this subsection, we extend our investigation to include the effects of $J_2$, which has comparable magnitude with $J_r$ derived from the expansion of Hubbard model, and explore the comprehensive $J_1$-$J_2$-$J_r$ phase diagram using the 36$A$ torus due to its respect for full lattice symmetry and its better capability in characterizing peaks in momentum space correlation functions. Extensive research has already been conducted on the ground state phase diagram of the $J_1$-$J_2$ model on the triangular lattice~\cite{Kaneko2014J1J2Tir,Hu2015J1J2Tri,Li2015J1J2Tri,Zhu2015J1J2Tri,Iqbal2016J1J2TriDSL,Saadatmand2016J1J2Tri,Ferrari2019J1J2Tri,Hu2019J1J2TriDSL,Tang2022J1J2TriExitedStateCross,Li2015YbMgGaO4PRL,Li2016J1J2Tri,Liu2016J1J2Tri,Luo2017J1J2Tri,Zhu2017J1J2Tri,Maksimov2019J1J2Tri,Wu2019J1J2Tri,Wu2021YbMgGaO4,Lauchili2024J1J2Tri}, providing a solid foundation for our current exploration. Numerous studies have shown that an increase in the next-nearest neighbor interaction leads to a transition from the 120$^{\circ}$ AFM order phase to a SL phase, followed by a transition into a collinear order phase. Based on the existing studies of the $J_1$-$J_2$ model, we introduce the ring exchange interaction and compute the results based on the 36$A$ torus. We obtain the $J_1$-$J_2$-$J_r$ ground state phase diagram presented in Fig.\ref{fig:PhaseDiagram} and identify ten distinct phases, which is quite rich. Magnetic ordered phases including the 120$^{\circ}$ AFM phase, collinear phase, zigzag phase and tetrahedral phase are detected. And there are also five nonmagnetic phases, like phase I-V. In addition, there is a finite region where there is spin liquid phase which is adiabatically connected to the SL in the $J_1$-$J_2$ model.

\begin{figure}[b]
	\centering
	\includegraphics[width=\linewidth]{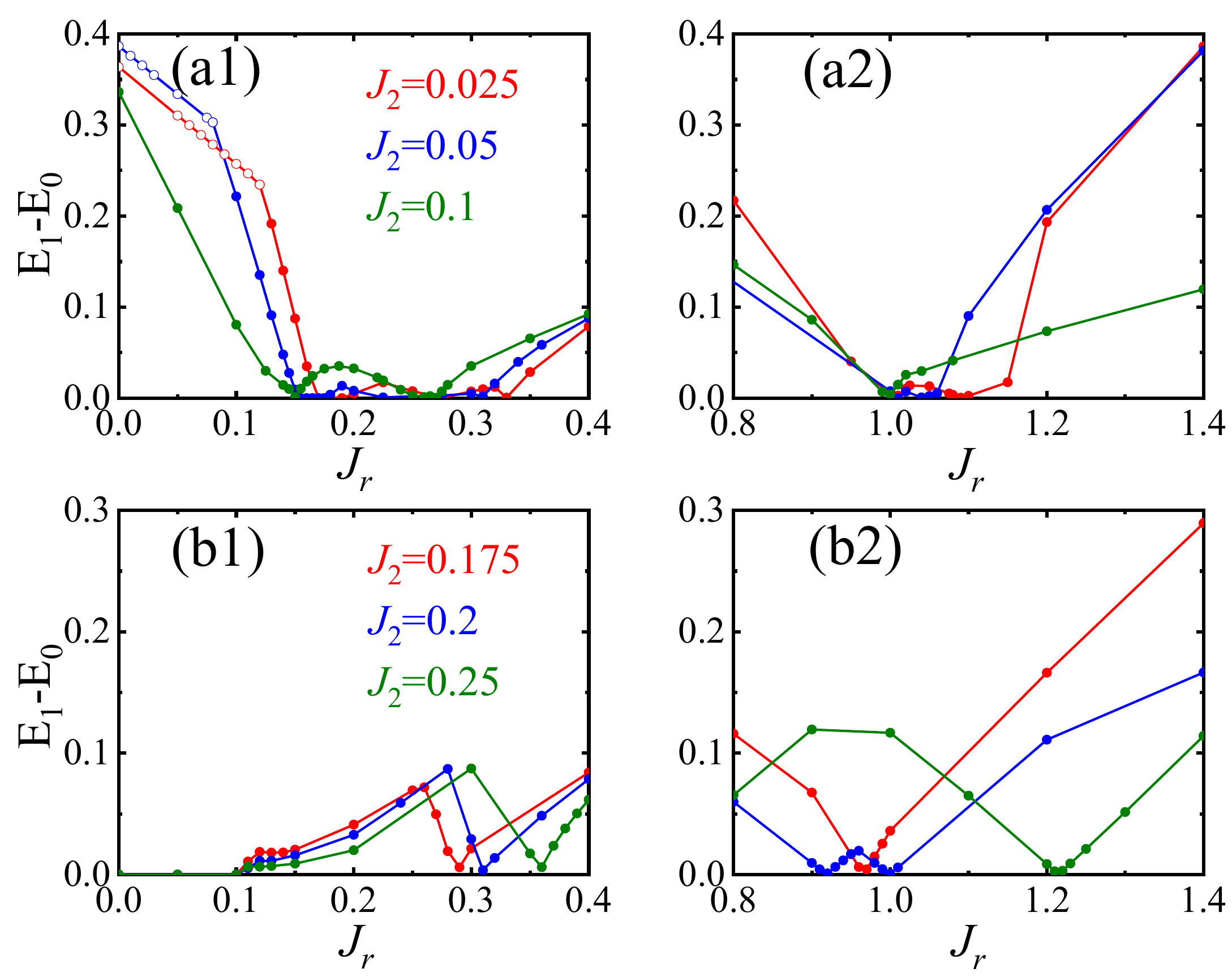}
        \caption{%
	The energy gaps E$_1-$E$_0$ of 36$A$ as functions of $J_r$. (a1-a2) $J_2=0.025,0.05,0.1$; (b1-b2) $J_2=0.175,0.2,0.25$. The solid and hollow circles represent the lowest singlet and triplet gaps, respectively.
  }
        \label{fig:E1-E0}
\end{figure}
 
 Figure~\ref{fig:E1-E0} shows the energy gap E$_1-$E$_0$ for different vertical paths in the $J_2$-$J_r$ phase diagram. When we fix $J_2$, as the parameters change, the ground state and the first excited state cross at different $J_r$ (see Fig.~\ref{fig:E1-E0}). Based on these crossing points, which indicate phase transitions, we obtain the phase diagram shown in Fig.~\ref{fig:PhaseDiagram}. When $J_2$ is relatively small, the phases induced by $J_r$ are similar to $J_2=0$ case. Take $J_2=0.05$ as an example, as shown in Fig.~\ref{fig:E1-E0}(a1), the lowest excited triplet state (hollow circles) intersects with the lowest excited singlet state at $J_r\approx0.08$, indicating a phase transition from the 120$^\circ$ AFM phase to a possible SL phase. As $J_r$ continues to increase, the ground state crosses with the first-excited singlet state at several different $J_r$. Similar to the $J_2=0$ case, we determine the phase transition points between the possible SL phase, phase I, phase II, zigzag, phase III, and phase IV by these level crossing points.
  \begin{figure}[t]
	\centering
	\includegraphics[width=\linewidth]{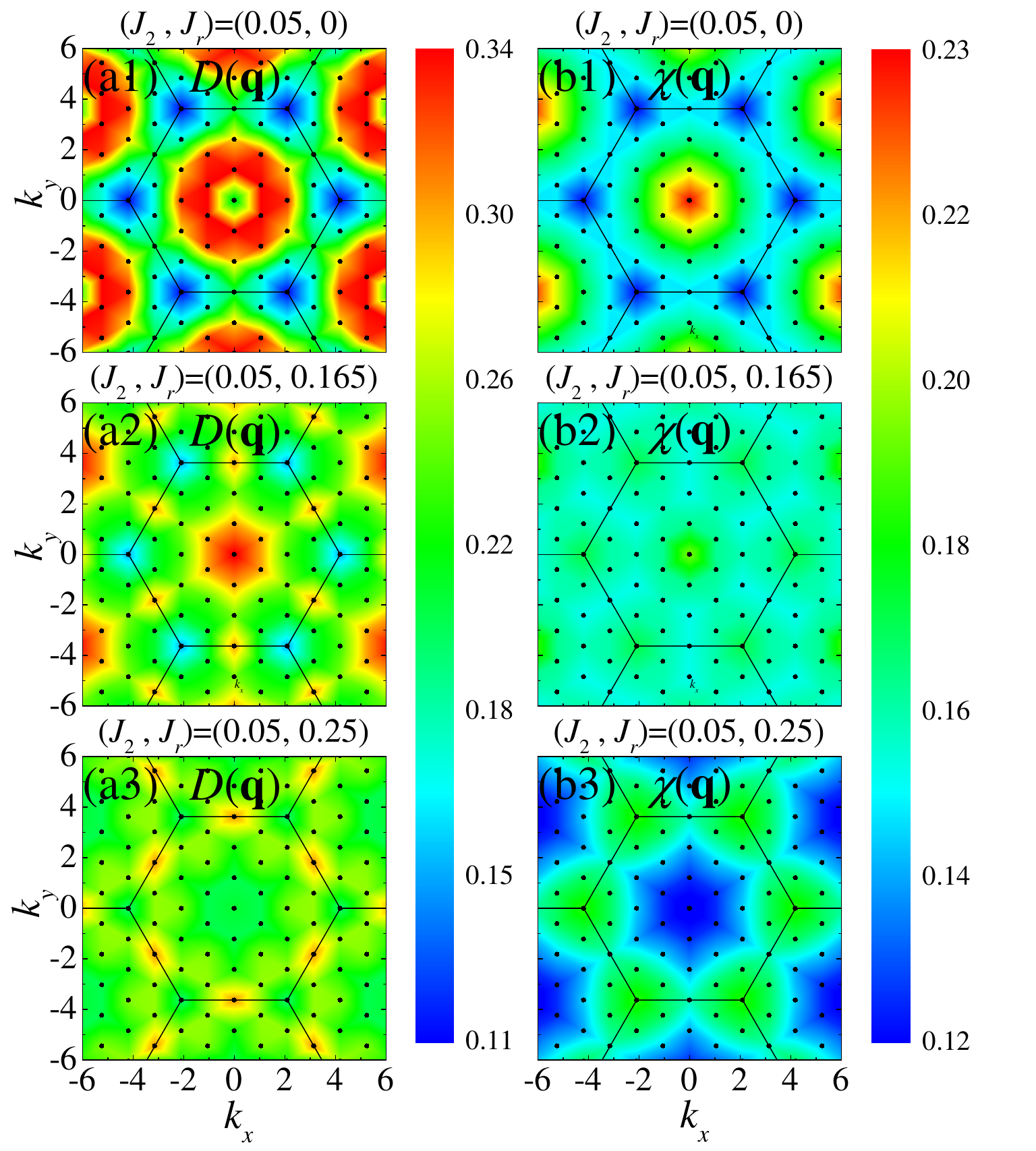}
        \caption{%
		(a1)-(a3) are dimer structure factors for 120$^{\circ}$ AFM phase, phase I and phase II at $J_2=0.05$, $J_r=0, 0.165$ and $0.25$, respectively. (b1)-(b3) are chiral structure factors for 120$^{\circ}$ AFM phase, phase I and phase II at $J_2=0.05$, $J_r=0, 0.165$ and $0.25$, respectively.
        }
        \label{fig:phase1+phase2}
\end{figure}
 From the structure factors presented in Fig.\ref{fig:phase1+phase2}, in phase I, we have found that adding a small $J_2$ leads to a decrease in the magnitude of $D(\mathbf{q})$ at the $M$ points, while it increases both $D(\mathbf{q})$ and $\chi(\mathbf{q})$ at the $\Gamma$ points, which means $J_2$ enhances scalar chiral correlation in phase I. However, it is still relatively weaker than $J_r=0$, $J_1$-$J_2$ case. While in phase II, the dimer and chiral correlations are nearly unaffected by $J_2$. Furthermore, the phase regions are also affected by the $J_2$ interaction. At $J_r\approx 0.2$, phase I occupies a very narrow region and will vanish as we increase $J_2$ to about $0.1$. For phase II, it extends to a wider region and eventually transitions to the tetrahedral phase at around $J_2\approx 0.13$. For phase III, the introduction of $J_2$ suppresses this phase, and the phase region shrinks to disappear at around $J_2\approx0.075$. For zigzag phase and phase IV, they extend to large areas, we do not see an obvious phase transition when increasing $J_2$ using the 36$A$ torus. For the SL phase in the $J_1$-$J_2$ triangular Heisenberg model, as $J_r$ increases, our ED results show that this SL phase expands to a wider region in the phase diagram. We compute the ground-state energy and its second derivative in the SL phase along the parameter path $J_r=0.17-1.36 \times J_2$ within the SL phase and find no signals of a phase transition. The approximate phase boundary between the 120$^{\circ}$ AFM phase and SL phase in Fig.~\ref{fig:PhaseDiagram} is determined by the intersections of low excited states which carry different spin quantum numbers~\cite{Tang2022J1J2TriExitedStateCross,Sandvik2016ExitedStateCross,Sandvik2018ExitedStateCross,Nomura2021ExitedStateCross,Sandvik2022ExitedStateCross,Lauchili2024J1J2Tri}, while other phase boundaries are obtained from level crossings between ground states and first excited states.

 We now focus on the large $J_2=0.25$ regime. At $J_r=0$, it is well studied that the collinear or stripe antiferromagnetic phase is the most favorable ground state. In this phase, magnetic moments are arranged parallel in one direction and anti-aligned in the other two directions on triangular lattice. The collinear phase exhibits three near-degenerate ground states $(\Gamma, 1)$ on the 36$A$ torus, where two states are completely degenerate and the other one is nearly degenerate. Therefore, we can see this degeneracy E$_1-$E$_0=0$ in Fig.~\ref{fig:E1-E0}(b1) when $J_r$ is small. 
  
\begin{figure}[t]
	\centering
	\includegraphics[width=0.9\linewidth]{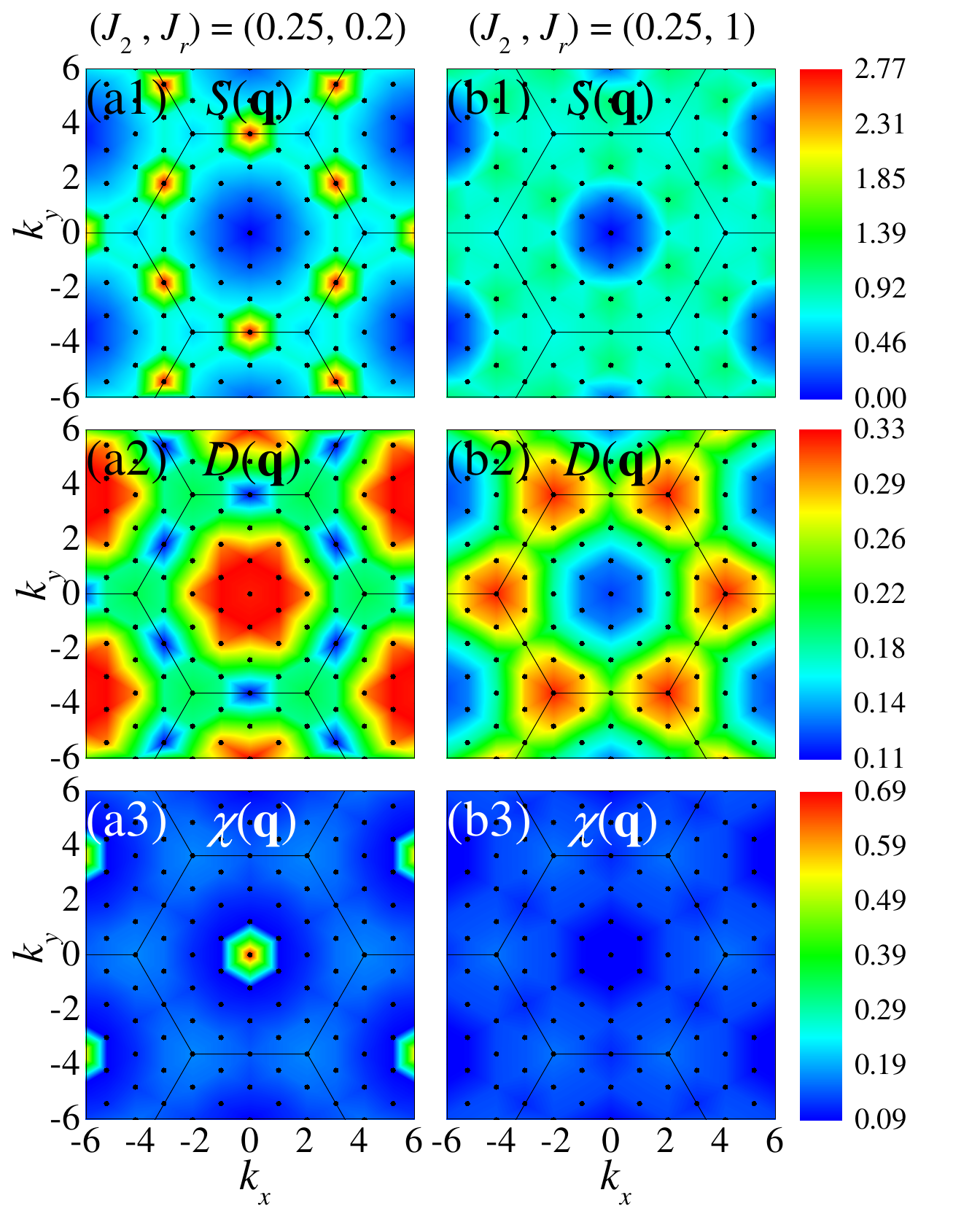}
        \caption{%
		(a1)-(a3) are structure factors for the tetrahedral phase at $J_2=0.25, J_r=0.2$ in 36$A$. (b1)-(b3) are structure factors for the phase V at $J_2=0.25, J_r=1$ in 36$A$.}
        \label{fig:Tetrahedral+phase5}
\end{figure}

 At $J_r\approx0.1$, the ground state transitions from two-fold singlets ($\Gamma, 1$) to a different one-fold singlet state ($\Gamma, 1$) [see Fig.~\ref{fig:E1-E0}(b1)]. The peaks of $S(\mathbf{q})$ remain at the $M$ points in this phase [see Fig.~\ref{fig:Tetrahedral+phase5}(a1)], which is similar to the collinear AFM phase. However, this phase has a strong scalar chiral correlation at $\Gamma$ point [see Fig.~\ref{fig:Tetrahedral+phase5}(c1)], which indicates a non-coplanar tetrahedral ordered phase where the spins point to the vertices of a regular tetrahedron. At relatively larger $J_r$, the system enters zigzag phase and phase IV. Between zigzag phase and phase IV, we find a new phase namely phase V. In this phase, the ground state is a singlet ($\Gamma, 1$). We observe no strong peaks in all structure factors [see Figs.~\ref{fig:Tetrahedral+phase5} (b1-b3)]. Consequently, it is a nonmagnetic state in this regime.
 
 We compute energy spectra for horizontal paths (keeping $J_r$ fixed while changing $J_2$) in these phases and find no energy level crossing between the ground state and the first excited state at $J_r$ = 0.5, 2.0 for 0$\leq J_2 \leq$ 0.25. We also calculate the second derivative of ground-state energy and find no signals of phase transition in the 36$A$ torus. These indicate that there is no extra phases in the entire zigzag phase and phase IV. These two phases at large $J_2$ are directly inherited from the $J_1$-$J_r$ model discussed before and have similar properties, respectively. For the 36$B$ torus, the zigzag phase regime is suppressed compared to 36$A$. And the magnetic Bragg peak are also found to be weaker after introducing $J_2$. That means the zigzag magnetic order is harder to be detected in the $J_2>0$ region.

We now compare our results with the $j_1$-$j_4$ model studied previously.
Ref.~\onlinecite{LiT2023RingVSED} studied the $j_1$-$j_4$ model through variational Monte Carlo (VMC), where $j_4\sum\limits_{[i,j,k,l]}(P_{i,j,k,l}+P_{i,j,k,l}^{-1})$ term contains four-site ring exchange coupling and other couplings. We expand the permutation operators in their model to spin operators, and analyze the relations between our model (with parameters $J_1$, $J_2$ and $J_r$) and theirs (with parameters $j_1$ and $j_4$), the results are $J_1=j_1+5j_4$, $J_2=j_4$ and $J_r=4j_4$~\cite{Xu2018RingDMRG,LiT2023RingVSED}. Based on these relations, the $j_1$-$j_4$ model with changing $j_4$ corresponds to $J_r=4J_2$ parameter path in our $J_2$-$J_r$ phase diagram (see the gray dashed line in Fig.~\ref{fig:PhaseDiagram}). In Ref.~\onlinecite{LiT2023RingVSED}, VMC identified a 120$^{\circ}$ AFM, two zigzag phases, and a VBS phase with $4\times6$ periodicity as $j_4$ increases, the number of phases and the phase boundaries are similar to our case. In our phase diagram, they correspond to 120$^\circ$ AFM, phase I, phase II and zigzag phase, respectively. Since the \(4\times 6\) state is incompatible with our 36-site clusters, we cannot identify this pattern. Ref.~\onlinecite{LiJX2023RingVMC} has also studied this model and found some other QSL phases in between 120$^{\circ}$ AFM and $U(1)$ SFS-QSL. These works inspire us to further investigate the ring exchange model using more unbiased and powerful methods such as $SU(2)$ DMRG and tensor network in the near future.

\section{Conclusion}\label{sec:Conclusion}

By employing Lanczos exact diagonalization (ED) techniques, we explore the interplay between the next-nearest-neighbor interaction $J_2$ and the four-spin ring-exchange $J_r$ in the triangular Heisenberg model. We perform ED calculations on two 36-sites tori to investigate ground-state properties and phase transitions. We derive a detailed $J_2$-$J_r$ ground-state phase diagram (see Fig.~\ref{fig:PhaseDiagram}). Upon introducing the $J_r$ and $J_2$ interaction, based on the correlation functions we computed, we observe several magnetic orderings, such as the tetrahedral and zigzag phases. Additionally, we discover several
nonmagnetic phases, labeled as phase I-V. The last three nonmagnetic phases are new phases that have not been studied before. We characterize all these phases to the best of our ability. Our findings align with some earlier studies, reveal some new results which can help to clarify the effects of ring exchange interaction on triangular lattice and guide to study using $SU(2)$ DMRG and tensor network method in the future. 

\begin{figure}[t]
	\centering
	\includegraphics[width=0.9\linewidth]{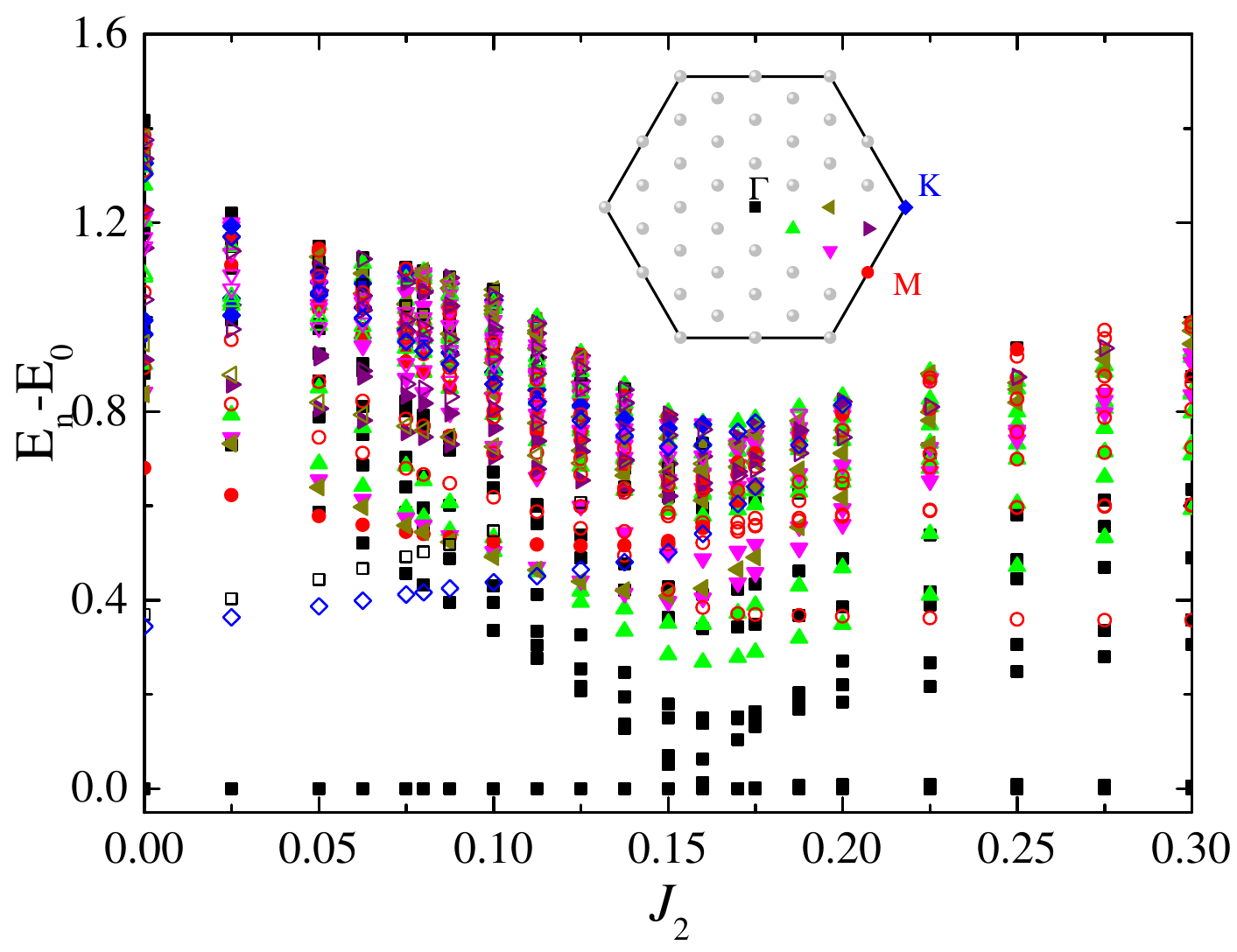}
        \caption{The energy spectrum of the $J_1$-$J_2$ Heisenberg model using the 36$A$ torus. The symbols with different shapes and colors represent the excited gaps $E_n-E_0$ obtained in different translational momentum sectors as shown in the inset. The solid and hollow points represent excited gaps with even and odd spin-inversion parity, respectively.}
        \label{fig:J1J2TriEngySpectrum}
\end{figure}

\begin{figure}[h]
	\centering
	\includegraphics[width=0.9\linewidth]{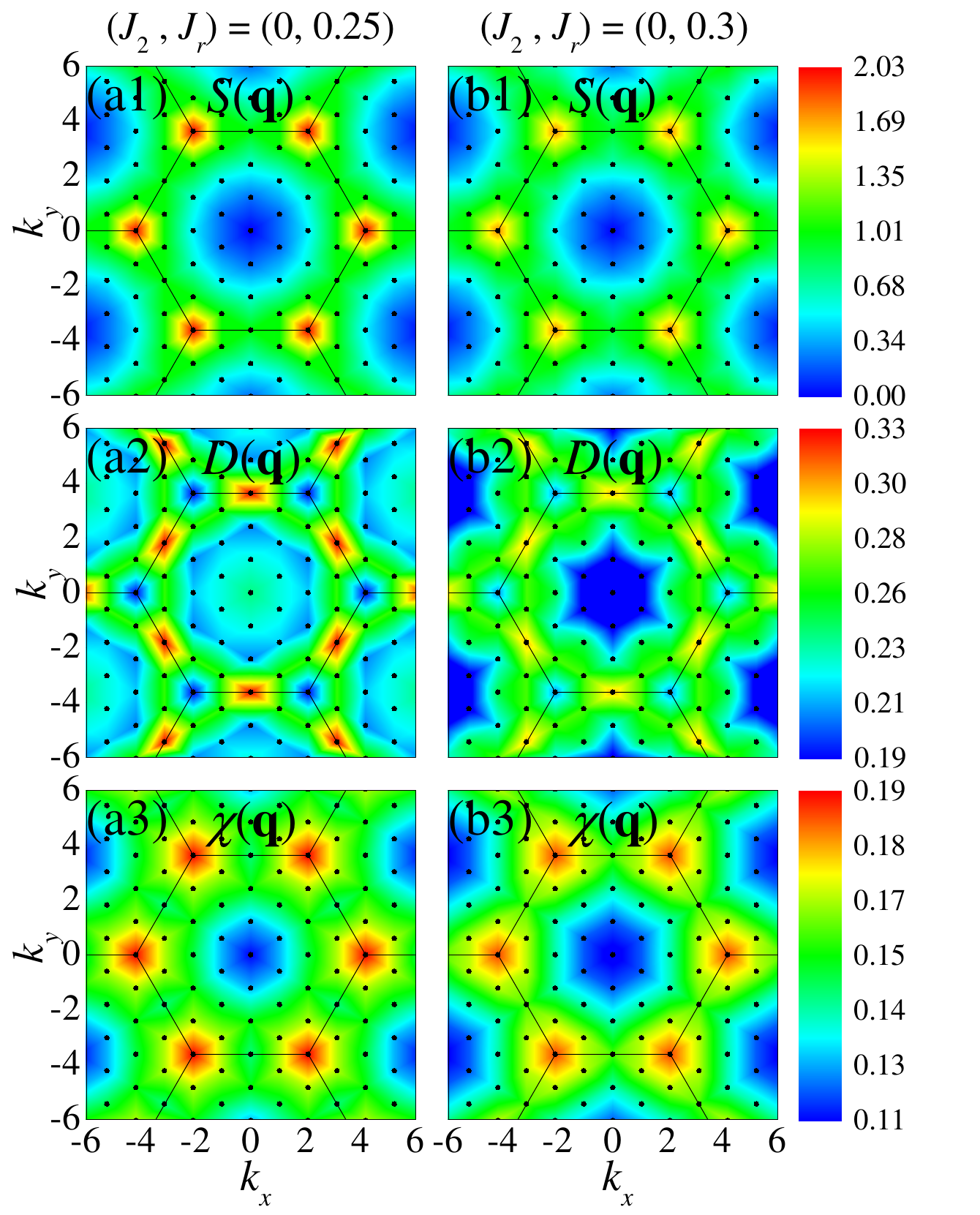}
        \caption{%
		The structure factors for phase II at $J_2=0, J_r=0.25$ and $J_2=0, J_r=0.3$ in 36$A$ torus, respectively.}
        \label{fig:phase2}
\end{figure}

\begin{acknowledgments}
This project is supported by Guangdong Basic and Applied Basic Research Foundation (Grant No. 2023B1515120013), NKRDPC-2022YFA1402802, NSFC-12474248, NSFC-92165204, NSFC-12494590, Leading Talent Program of Guangdong Special Projects (Grant No. 201626003), Guangdong Fundamental Research Center for Magnetoelectric Physics (Grant No.2024B0303390001), and Guangdong Provincial Quantum Science Strategic Initiative (Grant No.GDZX2401010).
\end{acknowledgments}

\appendix

\section{\texorpdfstring{Energy spectrum of the $J_1$-$J_2$ model}{}}
\label{App:J1J2}

Using the translational and spin inversion symmetry, as shown in Fig.~\ref{fig:J1J2TriEngySpectrum}, we obtain the energy spectrum of the $J_1$-$J_2$ Heisenberg model on triangular lattice and use different symbols to label the energy levels with different quantum numbers. When $J_2$ is small, the ground state is the 120$^\circ$ AFM phase and the lowest energy levels with different total spin $S$ form Anderson tower of states~\cite{Anderson1952TOS,Claire2005TOS,Wietek2017TOS}. With increasing $J_2$, there is an intersection between the lowest excited triplet [($K,1$), hollow blue diamonds] and the lowest excited singlet [($\Gamma,1$), solid black squares] occurring at $J_2\sim0.083$. Based on the previous studies with level spectroscopic method~\cite{Tang2022J1J2TriExitedStateCross,Sandvik2016ExitedStateCross,Sandvik2018ExitedStateCross,Nomura2021ExitedStateCross,Sandvik2022ExitedStateCross}, we identify this as the signal of the phase transition between the 120$^\circ$ AFM phase and the SL phase. When $J_2\gtrsim0.17$, three lowest singlet states ($\Gamma,1$) are quasi-degenerate, which indicates that the ground state turns into the collinear antiferromagnetic phase.

\section{\texorpdfstring{No phase transition at $Jr\approx 0.29$ in Phase II}{}}
\label{App:phase2}

As shown in Fig.~\ref{fig:EngySpectrumJP0p0}(a), an intersection between the first excited state and the ground state occurs at $Jr\approx 0.29$ in Phase II. To determine whether there is a phase transition here, we calculate the spin, dimer and chiral correlation at $J_2=0,J_r=0.25$ and $J_2=0,J_r=0.3$, which are shown in Fig.\ref{fig:phase2}. $S(\mathbf{q})$, $D(\mathbf{q})$ and $\chi(\mathbf{q})$ all show consistent peak positions with close magnitudes at these two sets of parameters. This indicates that the regime $0.22<J_r<0.29$ and $0.29<J_r<0.35$ should be in the same phase. And the level crossing at $Jr\approx 0.29$ does not correspond to a phase transition.

\section{\texorpdfstring{Dimer correlation of collinear AFM in $J_1$--$J_2$ model on square lattice}{}}
\label{App:SquareJ1J2}

\begin{figure}[h]
	\centering
	\includegraphics[width=\linewidth]{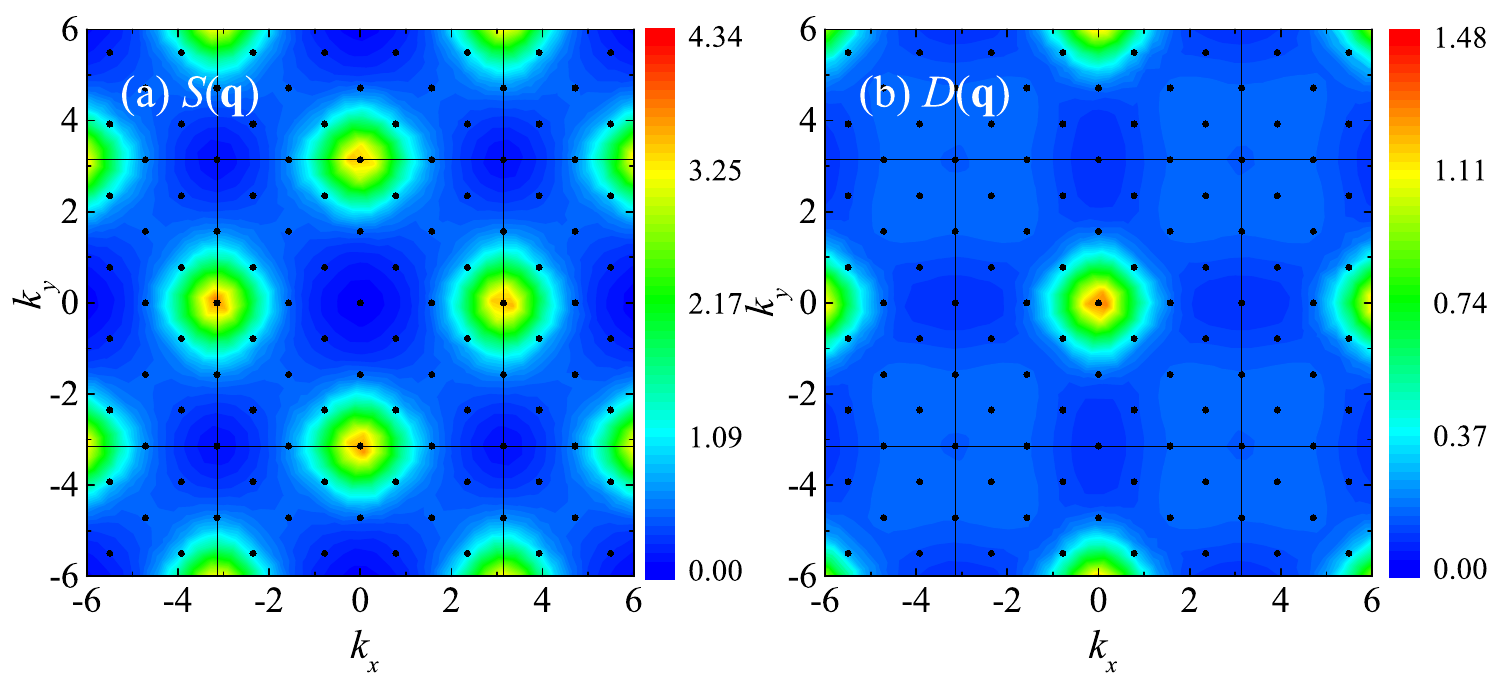}
        \caption{The structure factors for collinear AFM phase at $J_2=1.0$ in $J_1$--$J_2$ model on square lattice.}
        \label{fig:SquareJ1J2}
\end{figure}

For the $J_1$--$J_2$ model on square lattice, the ground state is the collinear AFM phase when $J_2$ is large. Figure~\ref{fig:SquareJ1J2} shows the spin and dimer structure factors obtained at $J_2$ = 1.0 on 32-site cluster. There is a strong peak located at $X$ points in the Brillouin zone, which is in agreement with the collinear AFM phase. And there is also a peak at $\Gamma$ points for the dimer structure factor shown in Fig.~\ref{fig:SquareJ1J2}(b). This peak is a manifestation of collinear magnetic order. And as shown in Fig.~\ref{fig:Corr-K}(e4), this phenomenon has also been seen in the zigzag phase of the $J_1$-$J_2$-$J_r$ Heisenberg model on the triangular lattice.

\bibliography{references}

\end{document}